\documentclass[11pt,a4paper]{article}
\usepackage{amsfonts}
\newcommand{\vsp}{$\vphantom{\Big |}$}
\newcommand{\hf}{{\textstyle\frac{1}{2}}}

\newcommand{\CC}{\mathbb{C}}
\newcommand{\ZZ}{\mathbb{Z}}

\newcommand{\bw}{{\textstyle\bigwedge}}
\newcommand{\Wp}{{\textstyle\bigwedge^+}}
\newcommand{\Wm}{{\textstyle\bigwedge^-}}
\newcommand{\DD}{I\kern-3.5pt D}
\newcommand{\FF}{I\kern-3.5pt F}
\newcommand{\arr}[1]{\smash{\mathop{\longrightarrow}\limits^{#1}}}
\def\one{{\mathchoice {\rm 1\mskip-4mu l} {\rm 1\mskip-4mu l}
        {\rm 1\mskip-4.5mu l} {\rm 1\mskip-5mu l}}}
\newcommand{\ah}{\theta(a)}
\newcommand{\bh}{\theta(b)}
\newcommand{\ch}{\theta(c)}
\newcommand{\tr}{\mbox{tr}\,}
\newcommand{\Tr}{\mbox{Tr}\,}
\newcommand{\Str}{\mbox{Str}\,}
\newcommand{\Tz}{\mbox{Tr}_z}

\begin{document}
\begin{center}
\LARGE\bf A Class of Anomaly-Free Gauge Theories
\end{center}
\vspace{3mm}
\begin{center}\large 
       G.\ Roepstorff\\
       Institute for Theoretical Physics\\
       RWTH Aachen\\
       D-52062 Aachen, Germany\\
       e-mail: roep@physik.rwth-aachen.de
\end{center}
\vspace{5mm}\par\noindent
{\bf Abstract}.
We report on a detailed calculation of the anomaly coefficients 
$\Tr(\ah\{\bh,\ch\})$ and $\Str(\ah\{\bh,\ch\})$ (trace and supertrace) for 
the reducible representation $\theta$ of a Lie algebra $\mbox{Lie}\,G$ on 
$\bw\CC^n$. Assuming that $G\subset U(n)$ where $n\ge2$, the representation $\theta$ is 
obtained from lifting the action of $U(n)$ on $\CC^n$ to the exterior algebra. 
The coefficients vanish provided $G\subset SU(n)$ and $n\ne3$. The singular
role of the group $SU(3)$ is emphasized.

\section{Introduction}

We recall that a gauge theory of massless fermions is said to be 
{\em chiral\/} if left- and
right-handed fermion fields transform differently under the gauge group $G$.
As is well known, this may cause a breakdown of classical symmetries on
the quantum level which manifests itself in the presence of {\em local
anomalies}, i.e., nonconservation of Noether currents. The abelian anomaly
has been discovered long time ago by Adler, Bell, and Jackiw, followed by
an explosion of the number of papers on the subject. For the early developments
see [1-7]. The problem has been reformulated over and over again. 
It is best understood using the Euclidean spacetime (compactified to $S^4$)
and functional integral methods [8,9]. The connection between anomalies and 
the Atiyah-Singer Index Theorem has noticed immediately [10-15]. Soon after,
anomalies were written on a BRS level in terms of differential forms.
An equation of constraint discovered by Wess and Zumino [16] defines
the anomaly in a direct way without recourse to a regularization scheme.
Mathematically speaking, the Wess-Zumino condition corresponds to a
cocycle condition in the affine space of gauge connections. For an account
of the history of the subject see the introductory chapter of the book
by R.A.\ Bertlmann [17].

Consistency of nonabelian chiral models requires that there be no local 
anomalies in 
the theory. Most models one might think of turn out to be inconsistent 
unless there is some group-theoretic reason for the anomalies to vanish.
For instance, one verifies consistency of the Standard Model 
by a routine calculation which is nothing but an exercise in (Lie) algebra.
The lesson of the Standard Model is that anomalies may cancel in
{\em reducible\/} representations of the gauge group even though the
irreducible constituents are anomalous. 
Recent results on chiral Schwinger models without gauge anomalies can be
found in [18] and applications to areas outside of particle physics appeared
in [19]. From the study of triangle diagram we
quote a general result: a chiral theory is free from anomalies in the
gauge currents if and only if some trace condition is satisfied involving 
the (represented) generators of the Lie algebra [20]. Granted this condition
all higher loop contributions vanish as well. The trace condition involves
the symmetrized third-order trace of the generators, called the {\em anomaly
coefficients}.

A Lie group is said to be {\em safe} if the anomaly coefficients vanish for
all its representations. Among the safe groups we find classical groups like
$SU(2)$, $SO(n)$ ($n\ne6$), and $Sp(2n)$ but also the exceptional groups
$G_2$, $F_4$, $E_6$, $E_7$, and $E_8$. Moreover, reducible real 
representations of nonsafe groups are anomaly-free. For instance, though 
the group $SU(3)$ is nonsafe, its representation $\{3\}\oplus\{\bar{3}\}$ is
real and thus has no anomalies. Similarly, the representation $\bw$ of
$SU(n)$ is real, hence anomaly-free for all $n$ (a special result of our
discussion in Section 4). None of these criteria, however, cover the case of 
the Standard Model. The famous cancellation of anomalies of leptons and quarks
is often seen as a miracle.

We will argue that this `miracle' in fact occurs in a large class
of reducible representations sharing two common features:
\begin{enumerate}
\item The gauge group $G$ is either $SU(n)$ ($n\ne3$) or a subgroup thereof.
\item Left-handed fermion fields transform according to the representation 
      $\bw^-$ of $G$ while right-handed fermion fields transform according 
      the representation $\bw^+$ (to be explained in the next section).
\end{enumerate}
To this we add the comment that the cancellation of anomalies fails if the
first condition is replaced by $G=U(n)$ and emphasize that the group $SU(3)$ 
has special features that prevent vanishing of the
anomaly coefficients in the representations $\bw^\pm$.
The Standard Model is now covered by the general result provided we
specialize it in the following way [21]:
\begin{enumerate}
\item The gauge group $G$ is a subgroup of $SU(5)$.
\item The Lie algebra Lie$\,G$ has the structure ${\bf su}(3)
\oplus{\bf su}(2)\oplus{\bf u}(1)$.
\end{enumerate}
The extra benefit of the present investigation is to learn that only the first 
condition is needed to effect the cancellation of anomalies.

We shall start considering the full unitary group $U(n)$ with $n\ge2$ 
and specialize to $SU(n)$ lateron.
The representations $\bw^\pm u$ of $u\in U(n)$ we focus on are very familiar 
constructions in linear algebra: they constitute the even and odd parts
of the representation $\bw u$ acting on the exterior algebra $\bw\CC^n$. 
As the argument presented below is purely algebraic (and to keep the paper 
short), we will refrain from discussing any aspects of particle physics in
relation to our result.

\section{The Exterior Algebra and $\ZZ_2$-Grading}

Since $U(n)$ is a classical group, it has a {\em defining representation\/}
given by the matrices $u\in U(n)$ viewed as linear operators on $\CC^n$.
Among many other representations
we single out those irreducible representations (irreps) that arise from 
lifting the defining represention to the exterior algebra $\bigwedge \CC^n$:
$$
        \matrix{\CC^n           &\arr{u}         & \CC^n          \cr
                \Big\downarrow  &                & \Big\downarrow \cr
                \bw\CC^n        & \arr{\bw u}    & \bw\CC^n       \cr}
$$
The representation thus obtained is denoted $\bw$. It has dimension $2^n$, 
is reducible, and may be decomposed into irreps $\bw^p$ acting on
$\bw^p\CC^n$ (the $p$th exterior power of $\CC^n$) of dimension 
$n\choose p$ in an obvious way:
\begin{equation}\label{list}
     \textstyle \bw =\bw^0\oplus\bw^1\oplus\cdots\oplus\bw^n\ .
\end{equation}
Another way of writing is
$$
        \bw^p u=u\wedge u\wedge\cdots\wedge u\qquad (p\ \mbox{factors}).
$$
The assumption 
is that, for $n$ appropriately chosen, no irreps other than those contained 
in the list (\ref{list}) are needed to accommodate the fundamental fermions  
encountered in reality. 

As soon as we confine ourselves to $SU(n)$, it is convenient to adopt yet 
another notation where each irrep is specified by its dimension $d$. However, 
if the irrep is complex, there are precisely two irreps of the same dimension:
given either one of them, its companion is obtained by complex conjugation.
In this case one writes $d$ and $\bar{d}$ to distinguish the two irreps. 
We may arrange all these irreps either in the diagram (varying $n$ but restricting to $n\le5$)
$$
  \begin{tabular}{|l|ccccccccccc|} \hline
$n$& \multicolumn{11}{c|}{irreps of SU(n)}\\ \hline
 2 &     &    &     &$\bw^0$&       &$\bw^1$ &    &$\bw^2$   &   &   &\vsp \\ 
 3 &     &    &$\bw^0$&    &$\bw^1$&    &   $\bw^2$&    &$\bw^3$&    &     \\
 4 &     &$\bw^0$&     &$\bw^1$&       &$\bw^2$&  &$\bw^3$&  &$\bw^4$&     \\
 5 &$\bw^0$&    &$\bw^1$&    &$\bw^2$&    &$\bw^3$&   &$\bw^4$&   &$\bw^5$ \\  
  \hline
  \end{tabular}
$$
or in a Pascal-like triangle indicating their dimensions:
$$
  \begin{tabular}{|l|ccccccccccc|} \hline
$n$ & \multicolumn{11}{c|}{irreps of SU(n)}\\ \hline
 2  &     &    &   & 1 &    & 2 &    & 1  &  \vsp        &    &            \\ 
 3  &     &    &  1 &    &  3   &    &   $\bar{3}$  &  & 1&      &         \\
 4  &     & 1 &     & 4 &       & 6 &                  &$\bar{4}$&    & 1 &\\
 5  & 1  &    & 5  &    & 10  &    & $\overline{10}$&  &$\bar{5}$&    & 1  \\  
  \hline
  \end{tabular}
$$
As for the group $SU(n)$, there is no distinction between the two 
representations $\bw^0$ and $\bw^n$. 
They are both one-dimensional and trivial. However, for $u\in$U(n) 
there is a distinction: $\bw^0u=1$ while $\bw^nu=\det u$.

As has been emphazised previously [21,22], the exterior algebra (as linear 
space) carries a $\ZZ_2$-graded structure making $\bw\CC^n$ a superspace:
$$
          \bw\CC^n=\bw^+\CC^n\oplus\bw^-\CC^n,\qquad
          \bw^+\CC^n=\sum_{p=\mbox{\scriptsize even}}\bw^p\CC^n,\quad
          \bw^-\CC^n=\sum_{p=\mbox{\scriptsize odd}}\bw^p\CC^n
$$
The representation $\bw$ of $U(n)$ respects the $\ZZ_2$-grading of $\bw\CC^n$
and decomposes as $\Wp\oplus\Wm$.  We may thus write
$$
      \bw u =\pmatrix{\bw^+u & 0\cr 0& \bw^-u\cr} ,\qquad u\in U(n)\ .
$$
Note that the dimensions of the even and odd subspaces are the same:
$$
              \mbox{dim}\,\bw^\pm\CC^n=2^{n-1}\ .
$$
From $\bw$ we construct the corresponding representation $a\mapsto\ah$ of the 
Lie algebra ${\bf u}(n)$ on $\bw\CC^n$: 
$$
   \ah =\frac{d}{dt}\bw\exp(ta)|_{t=0}
    =\pmatrix{\theta^+(a) &0\cr 0&\theta^-(a)\cr},\qquad
   \theta^\pm(a)\in{\rm End\,}\bw^\pm \CC^n\ .               
$$
The $\ZZ_2$-grading of the linear space $\bw\CC^n$ makes the endomorphism
algebra End$\bw\CC^n$ a {\em superalgebra}. See [22] for details. Since
the operator $\ah$ does not change the parity, it is said to be even or, 
to put it formally,
$$
         \ah\in\mbox{End}^+\bw\CC^n\ .
$$
Two types of traces are in use when dealing with superalgebras. There is
the ordinary trace, denoted Tr, and the supertrace, denoted Str.
The ordinary trace vanishes on commutators, while the supertrace vanishes
on supercommutators [23]. For the particular case at hand,
\begin{eqnarray}
     \Tr\bw u &=&\Tr\bw^+u+\Tr\bw^-u = \mbox{det}(1+u)   \label{trace}  \\
    \Str\bw u &=&\Tr\bw^+u-\Tr\bw^-u = \mbox{det}(1-u)   \label{super}
\end{eqnarray}
It is helpful to look at these formulas as obtained from a more general
trace evaluated at $z=\pm1$:
\begin{equation}
  \label{Ttz}
     \Tz \bw u = \sum_{p=0}^n z^p\,\Tr\bw^p u = \mbox{det}(1+zu)
   \qquad(z\in\CC)
\end{equation}
The formulas (\ref{trace}) and (\ref{super}) may be inverted to provide
those traces we are interested in:
\begin{equation}
  \label{pmin}
  \Tr\bw^\pm u =\hf(\Tr\bw u\pm\Str\bw u)\ .
\end{equation}
The ultimate goal is to compute traces of the form 
\begin{equation}
\label{tpm}
\Tr(\theta^\pm(a)\theta^\pm(b)\theta^\pm(c))=
\hf\Big(\Tr(\ah\bh\ch)\pm\Str(\ah\bh\ch)\Big)
\end{equation}
for $a,b,c\in{\bf u}(n)$ where $n\ge2$.
This task can now be reduced to computing the $z$-depending quantity 
$\Tz (\ah\bh\ch)$, referred to as the third-order trace.

\section{The Art of Computing Traces}

We continue to write 1 for group unit, but shall write $\bw1=\one$ for the
unit operator in End$\bw\CC^n$. The formula $\Tz \bw u=\mbox{det}(1+zu)$ can 
be rewritten as
\begin{equation}
  \label{Tt}
    \log\Tz \bw u=\tr\log(1+zu)\qquad u\in U(n),\ 1+zu\ne0.
\end{equation}
The simplest computation (taking $u=1$) leads to the zeroth-order trace:
\begin{equation}
  \label{zero}
                    \Tz \one=(1+z)^n
\end{equation}
A more involved problem is the computation of traces of order 1,2, and 3.
In a first step, we replace $u$ by $e^{ta}u$ in (\ref{Tt}) and take 
the derivative at $t=0$ to obtain
\begin{equation}
  \label{t1}
        \Tz (\ah\bw u)=z\,\Tz \bw u\ \tr(a\,u(1+zu)^{-1})\ .
\end{equation}
Hence, at the unit of the group, the result is a formula for the first-oder
trace: 
\begin{equation}
  \label{fir}
            \Tz \ah=z(1+z)^{n-1}\,\tr a\ .  
\end{equation}
In a second step, we
replace $u$ by $\exp(t\bh)u$ in (\ref{t1}) and again take the derivative at 
$t=0$:
\begin{eqnarray}
  \Tz (\ah\bh\bw u) &=& z\,\Tz (\bh\bw u)\ \tr(a\,u(1+zu)^{-1})\nonumber\\
                     &+& z\,\Tz (\bw u)\sum_{m=0}^\infty(-z)^m\sum_{k=0}^m
                         \tr\Big(a(\mbox{Ad}\,u)^kb\,u^{m+1}\Big)
                            \label{t2}
\end{eqnarray}
with $(\mbox{Ad}\,u)b=ubu^{-1}$, the adjoint representation. At the unit,
this creates a formula for the second-order trace:
\begin{equation}
  \label{sec}
  \Tz (\ah\bh)=z(1+z)^{n-2}(z\,\tr a\,\tr b+\tr(ab))\ .
\end{equation}
In a third step, we take $u=e^{tc}$ in (\ref{t2}) rewriting
\begin{equation}
  \label{spec}
z\sum_{m=0}^\infty(-z)^m\sum_{k=0}^m\tr\Big(a(\mbox{Ad}\,u)^kb\,u^{m+1}\Big)
=\sum_{n=0}^\infty t^n\,\tr\Big(a(\mbox{ad}\,c)^nb\,f_n(ze^{tc})\Big)
\end{equation}
where we made use of the Hausdorff formula
$$
 (\mbox{Ad}\,e^{tc})^kb=\sum_{n=0}^\infty \frac{k^n}{n!}t^n(\mbox{ad}\,c)^nb\ ,\qquad  (\mbox{ad}\,c)b=[c,b]
$$
and then introduced complex functions
$$
     f_n(z)=z\sum_{m=0}^\infty(-z)^m \sum_{k=0}^m\frac{k^n}{n!}
$$
which extend to analytic functions on $\CC\backslash\{-1\}$. Taking the 
derivative on both sides of (\ref{t2}) at $t=0$ when $u=e^{tc}$, we get a 
preliminary formula for the third-order trace:
\begin{eqnarray*}
  \Tz (\ah\bh\ch) &=&z(1+z)^{-1}\Tz (\bh\ch)\,\tr a   \\
                   & &+z(1+z)^{-2}\Tz \bh\ \tr(ac)        \\
                   & &+f_0(z)\,\Tz \ch\ \tr(ab)        \\
                   & &+zf'_0(z)\,\Tz \one\ \tr(abc)    \\
                   & &+f_1(z)\,\Tz \one\ \tr(a[c,b])
\end{eqnarray*}
It may now be shown that
$$
   f_0(z)=z(1+z)^{-2}\,\qquad f_1(z)=-z^2(1+z)^{-3}
$$
and thus 
\begin{eqnarray*}
  zf'_0(z)\,\Tz \one &=&z(1+z)^{n-3}(1-z)\\
    f_1(z)\,\Tz \one &=&-z^2(1+z)^{n-3}
\end{eqnarray*}
Putting all pieces of information together, we arrive at the final result
\begin{equation}
  \Tz (\ah\bh\ch)=(1+z)^{n-3}(z\alpha_1+z^2\alpha_2+z^3\alpha_3) \label{fin}
\end{equation}
where
\begin{eqnarray}
 \alpha_1 &=&\tr(abc)   \label{o1}\\
 \alpha_2 &=&\tr a\,\tr(bc)+\tr b\,\tr(ac)+\tr c\,\tr(ab)-\tr(acb)\label{o2}\\
 \alpha_3 &=&\tr a\,\tr a\,\tr c\ . 
\end{eqnarray}
If $n\ge 3$, the third-order trace comes out as an $n$th-order polynomial 
in $z$ as it should. For $n=2$, however, the formula (\ref{fin}) falsely
indicates the presence of singularity at $z=-1$ though we know in advance
that the trace ought to be a second-order polynomial. 
The solution to this discrepency is that, in two dimensions, there exist 
the identity
$$
   \tr(a\{b,c\})=\tr a\,\tr(bc)+\tr b\,\tr(ac)+\tr c\,\tr(ab)-\tr(abc)
   \qquad(n=2)
$$
so that
$$
 z\alpha_1+z^2\alpha_2+z^3\alpha_3=(1+z)(z\,\tr(abc)+z^2\,\tr a\,\tr b\,\tr c)
$$
and hence
\begin{equation}
  \label{ntwo}
       \Tz (\ah\bh\ch)=z\,\tr(abc)+z^2\,\tr a\,\tr b\,\tr c \qquad (n=2).   
\end{equation}
which is a much simpler expression that could also be derived by a
straightforward computation from scratch.

\section{Discussion of the Result}

We shall now apply the relations (\ref{fin}) and (\ref{ntwo}) obtained above 
to the cases of interest, i.e., when $z=\pm1$.
As a shorthand we introduce the following symmetric
functions
$$
  \alpha_\pm=\tr a\,\tr(bc)+\tr b\,\tr(ac)+\tr c\,\tr(ab)
              \pm\tr a\,\tr b\,\tr c\ .
$$
The ordinary trace, obtained when $z=1$, decomposes into a symmetric
and an antisymmetric contribution. As for the symmetric part, we
have the formula
\begin{equation}
  \label{sym1}
  \hf\Tr(\ah\{\bh,\ch\})=2^{n-3}\alpha_+  \qquad(n\ge2)  
\end{equation}
while the antisymmetric part reads:
\begin{equation}
  \label{ant1}
  \hf\Tr(\ah[\bh,\ch])= 2^{n-3}\,\tr(a[b,c])\qquad (n\ge2).  
\end{equation}
Next, we consider the case $z=-1$ in order to construct the supertrace 
which again decomposes into (anti)symmetric contributions. The symmetric 
part is given by
\begin{equation}
  \label{sym2}
  \hf\Str(\ah\{\bh,\ch\})
                   =\cases{\tr a\,\tr b\,\tr c-2^{-1}\alpha_- & if $n=2$  \cr
                                 \alpha_--\tr(a\{b,c\})       & if $n=3$  \cr
                                    0                         & if $n\ge4$\cr}
\end{equation}
while the antisymmetric part reads:
\begin{equation}
  \label{ant2}
  \hf\Str(\ah[\bh,\ch])=\cases{-2^{-1}\tr(a[b,c]) & if $n=2$  \cr
                                      0           & if $n\ge3$\cr}  
\end{equation}
The left-hand side of (\ref{ant1}) does not vanish unless the Lie algebra
consists of trace-less matrices. Therefore, it is not conceivable that
the anomaly coefficients vanish unless $\tr a=\,\tr b=\,\tr b=0$ which is
what we shall assume from now on. In effect, we are dealing then with
gauge groups $SU(n)$ or with subgroups thereof.

With vanishing traces, formulas simplify considerably. We particularly
obtain the following result for the anomaly coefficients in the
representations $\bw^\pm$:
\begin{equation}
  \label{sun}
  \Tr(\theta^\pm(a)\{\theta^\pm(b),\theta^\pm(c)\})=
  \cases{0                  & if $n=2$ or $n\ge4$  \cr
         \mp\,\tr(a\{b,c\}) & if $n=3$             \cr}
\end{equation}
The coefficients vanish in any dimension except when $n=3$. It is perhaps
surprising that gauge theories based on $SU(3)$ play a distinguished role.

For completeness
we mention the result for the corresponding symmetric coefficients:
\begin{equation}
\label{antn}
  \Tr(\theta^\pm(a)[\theta^\pm(b),\theta^\pm(c)])=
  \cases{2^{-1}(1\mp1)\,\tr(a[b,c])      & if $n=2$   \cr
               2^{n-3} \,\tr(a[b,c])      & if $n\ge3$ \cr}
\end{equation}
Note that $\tr(a[b,c])$ are presicely the structure constants of the Lie
algebra. The relations (\ref{ant1}) and (\ref{antn}) confirm the expectation
that the structure constants come out the same in any faithful representation,
apart from some natural number in front.

Nowhere in the calculation have we used the assumption the group elements
$u$ are unitary. Nor have we used the relation $a^*=-a$ for the elements
$a$ of the Lie algebra. Hence our results hold equally well when
the unitary group $U(n)$ is replaced by the full linear group $GL(n,\CC)$
and $SU(n)$ is replaced by the unimodular group $SL(n,\CC)$. However,
noncompact groups are not favoured as candidates for symmetries in particle
physics.

\vspace{2cm}
{\Large\bf References}
\vspace{3mm}
\begin{enumerate}
%1-5:
\item S.L.\ Adler: Phys.Rev.\ {\bf 177} (1969) 2426
\item J.S.\ Bell and R.\ Jackiw: Nouvo Cim.\ A {\bf 60} (1969) 47
\item W.A.\ Bardeen: Phys.Rev.\ {\bf 184} (1969) 1848
\item S.L.\ Adler: {\em Lectures on Elementary Particles and Quantum Field
      Theory}, Ed.\ Deser, MIT, Cambridge, MA 1970
\item S.\ Coleman and B.\ Grossman: Nucl.Phys.\ B {\bf 203} (1982) 205
%6-10:
\item L.\ Baulieu: Nucl.Phys.\ B {\bf 241} (1884) 557
\item R.\ Stora: {\em Progress in Gauge Field Theory}, (Cargese 1983)
      Plenum New York 1984
\item K.\ Fujikawa: Phys.Rev.\ D {\bf 21} (1980) 2848, Erratum: 
      D {\bf 22} (1980) 1499
\item K.\ Fujikawa: Phys.Rev.\ D {\bf 25} (1982) 2584
\item R.\ Jackiw and C.\ Rebbi: Phys.Rev.\ D {\bf 14} (1976) 517
%11-15:
\item R.\ Jackiw and C.\ Rebbi: Phys.Rev.\ D {\bf 16} (1977) 1052
\item N.K.\ Nielsen and B.\ Schroer: Nucl.Phys.\ B {\bf 127} (1977) 493
\item N.K.\ Nielsen, H.R\"omer, and B.\ Schroer: Phys.Lett.\ {\bf 70B} (1977)
      445
\item L.\ Alvarez-Gaum\'e and P.\ Ginsparg: Ann.Phys.\ {\bf 161} (1985) 423
\item L.\ Alvarez-Gaum\'e and P.\ Ginsparg: Nucl.Phys.\ B {\bf 243} (1984) 449
%16-20:
\item J.\ Wess and B.\ Zumino: Phys.Lett.\ {\bf 37B} (1971) 95
\item R.A.\ Bertlmann: {\em Anomalies in Quantum Field Theory}, Oxford Science
      Publications, Clarendon Press Oxford (UK) 1996
\item H.\ Grosse and E.\ Langmann: hep-th/0004176
\item J.\ Fr\"ohlich and B.\ Pedrini: hep-th/0002195
\item R.\ Ticciati,{\em Quantum Field Theory For Mathematicians},
      Encyclopedia of Mathematics and its Applications {\bf 72},
      Cambridge University Press 1999
%21-23:
\item G.\ Roepstorff: hep-th/9907221
\item G.\ Roepstorff and Ch.\ Vehns: math-ph/9908029
\item N.\ Berline, E.\ Getzler, and M.\ Vergne, {\em Heat Kernels and
      Dirac Operators}, Springer Berlin Heidelberg 1992
\end{enumerate}
\end{document}